\documentclass[10pt,twocolumn,prd,aps,amssymb,amsmath,tightenlines]{revtex4}
\usepackage{psfig}
\usepackage{graphics}
\usepackage[pdftex]{graphicx}

\begin{document}
\preprint{}

\title{
Implications of the L\"uders Postulate for Quantum Algorithms
}
\author{Bernhard K. Meister}

\affiliation{
 Physics Department, Renmin University of China, Beijing 100872,
 China}
\date{\today}

\begin{abstract}
The L\"uders postulate is reviewed and implications for quantum
algorithms are discussed. A search algorithm for an unstructured
database is described.
%

\end{abstract}

\maketitle

\section{Introduction}

In this paper the L\"uders postulate\cite{luders} 
is used to develop a new quantum computer algorithm to  search an
unstructured database
 with exactly one {\sl marked} record.
 The L\"uders postulate was introduced as a modification of the original
measurement theory of quantum
mechanics as presented by von Neumann\cite{neumann}. 
   It describes unambiguously the measurement
   process
   of observables
with a degenerate spectrum.
Our approach to the unstructured
 database search
 differs from
 Grover's\cite{grover} original algorithm due to our focus on the details of the measurement process.
This paper emphasizes conceptual issues, while implementability
will be addressed in detail elsewhere.

 The paper is organised as follows. In  Section~\ref{sec:2} some notational
 issues and other preliminaries are covered.
 In Section~\ref{sec:3} the L\"uders postulate is reviewed and with the
 help of a simple example some implications of the postulate are clarified.
In Section~\ref{sec:4} a unstructured database search algorithm is
described.
In the last section some concluding remarks are added.



\section{Preliminaries }
\label{sec:2}

Let us begin by presenting the relevant notation.
We define, following the conventions of the field, some terms. A
{\sl qubit} represents the superposition of a pair of orthogonal
quantum states. The pair of states will be denoted as $|0\rangle$
and $|1\rangle$. 
Implementations of qubits
 have been discussed in various settings, such as
spin-$\frac{1}{2}$ particles, polarised photons, {\it et cetera}.
For reasons of simplicity we assume that the qubits are either
implemented as spin-$\frac{1}{2}$ particles or as polarised
photons. Other potential implementations of qubits are equally
valid and do not change the basic nature of the algorithm.
Spin-up and spin-down particles, or horizontally and vertically
polarised photons, eigenstates with respect to a particular
direction, are written respectively as $|0\rangle$ and
$|1\rangle$.

Extensions of single qubits are obtained as usual by constructing
a multi-particle tensor product. The basis of the $n$ qubit states
can be written in the form: $|e_{n}\rangle \otimes |e_{n-1}\rangle
\otimes \cdots \otimes |e_1\rangle$, where each $|e_i\rangle$ is
either $|0\rangle$ or $|1\rangle$. For reasons of brevity we often
write $|e_{n} e_{n-1} \cdots e_1\rangle$ for the $n$ qubit. We can
also write qubits in vector notation, where each of the $N$ states
$|000\cdots00\rangle$ to $|111\cdots11\rangle$ corresponds to a
vector $\vec{e}_j$ of length $N$ containing one nonzero component
of value $1$ in the $j$-th place.

The following additional notation will also be used. The records
of a database with $N=2^{n+1}$ elements are encoded in the set
$D=\{|\omega_i\rangle\}$ $(i=0,1,2,...,N-1)$, where
$|\omega_i\rangle$\footnote{One can in the same way encode the
possible solutions of a NP-complete problem like $3-SAT$.} is
given by the binary representation of length $n+1$ of the numbers
$0$ to $N-1$. Therefore, each state $|\omega_i\rangle$ is
represented by a particular sequence $|e_{n} e_{n-1} \cdots
e_{1}\rangle$, where each $e_i$ is either $0$ or $1$.
The marked record, in particular, will be denoted
$|\omega_k\rangle$. Normalisation factors associated with
combination of states\footnote{In general, but not with absolute
consistency, we call a {\sl state} any part of a quantum
mechanical wave function that can be written in terms of exactly
one of the basis elements $|e_{n} e_{n-1} \cdots e_{1}\rangle$.}
are consistently ignored until the end of Section IV.

\section{ The L\"uders Postulate}
\label{sec:3}

The L\"uders postulate\cite{luders} \& \cite{isham} describes the
measurement process of observables with a degenerate spectrum, and
it has become part of the standard canon of quantum mechanics. In
the case of operators with a degenerate spectrum it postulates
that the projection of the initial wave function is onto exactly
one point in each degenerate subspace. The point chosen is the
element of the degenerate subspace `closest' - in terms of
transition probability - to the initial wave function.
This `refinement' of von Neumann's projection postulates\footnote{
In von Neumann's original approach the initial wave function is
projected onto a full basis, where the choice of basis depends for
the degenerate subspaces on the exact nature of the measurement
apparatus.} seems reasonable, since L\"uders' postulate produces
measurements that disturb the wave function minimally.

The mathematical formulation of the postulate is given next using
standard Dirac notation. We define the normalized eigenfunctions
of the observable $\hat{O}$ with $K$ different eigenvalues, each
having the degeneracy $d_k$, to be
\begin{eqnarray}
|\psi_{k,j}\rangle ,
\end{eqnarray}
where
 $k=1,2,...,K$
and $j=1,2,...,d_k$. The eigenfunctions allow the definition of
the following set of $K$ projection operators\footnote{Any choice
of basis of the degenerate subspace leads to the same projection
operator.}
\begin{eqnarray}
\hat{P}_k=\sum_{j=1}^{d_k}|\psi_{k,j}\rangle \langle \psi_{k,j}| .
\end{eqnarray}

A measurement of an arbitrary pure state
$|\phi\rangle$\footnote{The application of L\"uders' postulate to
entangled states is also of interest.} now gives according to the
L\"uders postulate the `reduction' to the following states
\begin{eqnarray}
|\phi\rangle \rightarrow {\rm Prob}[O=\lambda_k]^{-1/2} \hat{P}_k
|\phi\rangle
\end{eqnarray}
with the probabilities for the distinct eigenvalues of
\begin{eqnarray}
{\rm Prob}[O=\lambda_k]=\langle \phi | \hat{P}_k|\phi\rangle .
\end{eqnarray}

The impact of the L\"uders postulate on the distinguishability of
similar observables is next presented. The aim is to distinguish
two known observables with the help of a measurement of an input
wave function. We consider observables that measure individual
spin-$\frac{1}{2}$ particles.  The observable $\hat{O}$ is either
chosen to be the identity operator
\begin{eqnarray}
\hat{I} =
\left(
\begin{array}{lr}
1 &  0  \\
0 & 1
\end{array} \right),
\end{eqnarray}
or the operator
\begin{eqnarray}
\hat{J} = 
\left(
\begin{array}{lc}
1 &  0 \\
0 & 1+\delta
\end{array} \right)
\end{eqnarray}
that associates the eigenvalue $1$ to the eigenstate spin-up and
$1+\delta$ to the eigenstate spin-down.

 A measurement of either the observable $\hat{I}$ or $\hat{J}$ for an
 input wave function in equal superposition of
spin-up and spin-down, i.e. $1/\sqrt{2}(|1\rangle+ |0\rangle)$, is
carried out next. It gives for the first observable $\hat{I}$ a
direct projection of the wave function onto itself. For the second
observable $\hat{J}$ the measurement outcome is a mixed state with
equal probability in the state spin-up and spin-down
as long as $\delta$ is nonzero.

The unique outcome $1/\sqrt{2}(|1\rangle+ |0\rangle)$ for the
first observable can be distinguished simply from the mixed state
outcome by standard interference techniques. One can for example,
in an additional apparatus, measure the probability of the wave
function in an appropriate basis like $1/\sqrt{2}(|1\rangle+
|0\rangle)$ and $1/\sqrt{2}(|1\rangle - |0\rangle)$. In the first
case the outcome will always be $1/\sqrt{2}(|1\rangle+
|0\rangle)$. For the second case, the mixed state,
 the
probability for each of the basis states is $1/2$. Therefore, the
ability to distinguish the two observables below any chosen error
threshold $\epsilon$ is possible, if sufficient identical copies
of the system are prepared. Each copy available decreases the
probability of a mistake by a factor of $1/2$. If $m$ copies are
prepared, the probability of an incorrect choice is $2^{-m}$.

 In effect, this simple example already
captures the essence of the paper. Namely, {\it an infinitesimal
deformation of an observable  changes a degenerate into
nondegenerate spectrum and can lead, as in the case described, to
an observable difference}.

In the coming section we link the form of the observable through
the use of an oracle to the location of the {\sl marked} state.
It will turn out that these different observables, i.e. different
locations of the {\sl marked} state, can be distinguished
efficiently with the help of measurements of specially prepared
wave functions.
 As an aside, the space of Hermitian operators possesses a natural Finslerian
metric\cite{finsler} permitting a more comprehensive study of
their properties.


\section{Sketch of the Algorithm }
\label{sec:4}

After having introduced the necessary notation and the underlying
principle behind the algorithm in the previous two sections, we
can describe the process for finding the {\sl marked} state in an
unstructured database.

Let us begin by presenting an outline of the algorithm.
The algorithm divides the database search into smaller pieces to
make it more tractable. We do so, initially, by dividing the whole
set of records into two equal subsets. The algorithm determines
within certain error bounds, which one of the two subsets contains
the {\sl marked} state. This is done by linking the form of the
observable to the location of the {\sl marked} state. The
different observables, i.e. different locations of the {\sl
marked} state, are used to `measure' (following L\"uders'
postulate) specially prepared wave functions. This measurement,
i.e. `collapse' into the relevant eigenstates, gives us
information to determine (with the help of a simple further
measurement) up to a certain accuracy the presence or absence of
the {\sl marked} state in the subset under consideration.
Repetitions of the process decrease the error
probability. 
Once the set containing the {\sl marked} state has been identified
to a sufficient level of accuracy, we further divide the set into
two subsets of equal size and restart the process.
Without loss of generality we may set $N=2^{n+1}$. The division of
the records into subsets has to be carried out $n+1$ times, until
we are finally left with the unique marked record.  We define a
{\sl cycle} to be the process of halving the number of states
under consideration.
This section describes the first {\sl cycle} of the algorithm.

We present in the following three elements of the algorithm. The
wave function to be measured is constructed first. A description
of the possible observables comes second. Finally, we let the
observable act on the wave function.

\noindent {\bf A. Wave function}. We begin by transforming the $n$
qubit wave function of the form $|000\cdots00\rangle$ into the
`input' wave function that will be measured by the observable.
This is done by creating the superposition of all $2^n$ states of
the form $|000\cdots00\rangle + |000\cdots01\rangle +
|000\cdots10\rangle+ \cdots + |111\cdots11\rangle$ from the
starting wave function 
by a sequence of
Walsh-Hadamard transforms on the $n$ individual qubits\footnote{We
initially rotate the first qubit to produce $|000\cdots01\rangle +
|000\cdots00\rangle$ and then step by step rotate the other
qubits. This produces the desired wave function.}. As noted above,
we shall ignore for simplicity the normalization factors. This
will not affect the argument.

\noindent {\bf B. Observable}. Next, we construct an observable
that depends on the position of the {\sl marked} record.
It will turn out to be the symmetrised product of two matrices. 
One matrix is a fixed Hermitian matrix containing two block
diagonal submatrices with the associated real, nonzero and unequal
eigenvalues $a_1$ and $a_2$ (e.g. $a_1=1+\delta$ and $a_2=1$),
where the first subspace is $2$-dimensional. Two eigenvectors,
which span the $2$-dimensional subspace of the first submatrix,
are chosen to be the sum of all the individual states,
$|000\cdots00\rangle + |000\cdots01\rangle + |000\cdots10\rangle+
\cdots + |111\cdots11\rangle$, and the sum of all the states with
alternating sign, $|000\cdots00\rangle - |000\cdots 01\rangle+
|000\cdots 10\rangle- |000\cdots 11\rangle
\cdots+|111\cdots10\rangle-|111\cdots11\rangle$.
In matrix form
this can be written, due to the Schur decomposition, as the
product of three matrices of the form $\hat{A}=
\hat{R}^{\dagger}\hat{G} \hat{R}$ with $\hat{G}$ equal to

\begin{eqnarray}  
\left(
\begin{array}{lccccccr}
a_1 &  0 & 0&0&\cdots&0&0&0 \\
0 & a_1 & 0&0&\cdots&0&0&0\\
0 & 0 & a_2&0&\cdots&0&0&0\\
0 & 0 & 0&a_2&\cdots&0&0&0\\
\vdots&\vdots&\vdots&\vdots&\ddots&\vdots&\vdots&\vdots\\
0 & 0 & 0&0&\cdots&0&a_2&0\\
0 & 0 & 0&0&\cdots&0&0&a_2\\
\end{array} \right),
\end{eqnarray}
and $\hat{R}$ a simple unitary basis rotation matrix transforming
the initial basis $\vec{e}_1,\vec{e}_2,...,\vec{e}_N$ into the new
basis with the first two basis elements $|000\cdots00\rangle +
|000\cdots01\rangle + |000\cdots10\rangle+ \cdots +
|111\cdots11\rangle$ and $|000\cdots00\rangle - |000\cdots
01\rangle+ |000\cdots 10\rangle- |000\cdots 11\rangle
\cdots+|111\cdots10\rangle-|111\cdots11\rangle$, which are
identical to the two basis elements described above. The rest of
the new basis elements that make up the rows of $\hat{R}$ can be
chosen arbitrarily as long as the resulting matrix is unitary. The
product of the three matrices $\hat{R}^{\dagger}\hat{G} \hat{R}$
is Hermitian.

The other matrix $\hat{B}$ associates a phase of $e^{i\pi}$, if
the state is the marked state and otherwise leaves the state
unchanged. This operation  is the standard delta-function oracle
of the form $f_k(\omega_i)=\delta_{ik}$, implemented in the
quantum mechanical context as
\begin{eqnarray}
|\omega_i\rangle  \to  (-1)^{\delta_{ki}}|\omega_i\rangle ,
\end{eqnarray}
where $|\omega_k\rangle$ is the marked state. Note that this
transformation is identical to the conventional oracle employed in
the unstructured database search algorithm\footnote{In the case of
NP-complete problems, ala 3-SAT, instead of an oracle, one can
associate the phase change to the solutions of the problem in
hand. Each of the $2^N$ possible solutions has either its phase
unchanged or gets a phase change of $e^{i\pi}$.}.
Feasibility and other issues related to oracles have been
discussed in the literature -
 see Nielsen and Chuang \cite{chung} for details and references.


Each of the two operators $\hat{A}$ and $\hat{B}$ is Hermitian on
its own, but the product might not, because the matrices do not
necessarily commute. Instead we create a Hermitian observable by
symmetrizing the product
\begin{eqnarray}
\hat{C}=(\hat{A} \hat{B} + \hat{B} \hat{A})/2.
\end{eqnarray}
The observable $\hat{C}$ will be used in the measurement process.

\noindent {\bf C. Measurement}. We carry out the necessary
measurements to distinguish between the different possible
observables with sufficient accuracy. We let the observable
$\hat{C}$ act on the `input' wave function $|000\cdots00\rangle +
|000\cdots01\rangle + |000\cdots10\rangle+ \cdots +
|111\cdots11\rangle$. There are two cases to be studied. In the
first case there is no solution in the subset considered. The
`input' wave function, i.e. an eigenstate of the first subspace of
$\hat{A}$, is then also the `output' wave function. The operators
$\hat{A}$ and $\hat{C}$ are identical and the result of the
measurement is a pure state.

The case where there is a solution in the subset considered is
more interesting to analyze.
Let us define the vectors $\vec{u}_1$ and $\vec{u}_2$ to be a
basis of the two-dimensional degenerate subspace of $\hat{A}$.
They are chosen to be of the form $|000\cdots00\rangle +
|000\cdots10\rangle + \cdots +|111\cdots00\rangle+
|111\cdots10\rangle$, i.e. the sum of only the `even' states, and
$|000\cdots01\rangle + |000\cdots11\rangle + \cdots
+|111\cdots01\rangle+ |111\cdots11\rangle$, i.e. the sum of only
the `odd' states. For the {\sl marked} state in any {\sl odd}
position we have $\hat{C} \vec{u}_1= a_1 \vec{u}_1$ and $\hat{C}
\vec{u}_2 = a_1 \vec{u}_2 + \vec{v}_2$ with $\vec{v}_2$ a nonzero
vector and element of the larger degenerate subspace of $\hat{A}$.
For the {\sl marked} state in any {\sl even} position we have
$\hat{C} \vec{u}_2= a_1 \vec{u}_2$ and $\hat{C} \vec{u}_1 = a_1
\vec{u}_1 + \vec{v}_1$ with $\vec{v}_1$ similar to
$\vec{v}_2$\footnote{ Without loss of generality we set the marked
state $\vec{e}_k$ to be the $k$-th record, i.e. $\vec{e}_k$, and
equal to $\vec{u}+\vec{v}$, with $\vec{u}$ and $\vec{v}$
eigenstates of the first and second degenerate subspace of
$\hat{A}$ respectively. Simple linear algebra shows that
$\hat{C}\vec{u}= d_1 \vec{u}+d_2 \vec{v}$ with $d_1=a_1(1+u')$ and
$d_2=\frac{1}{2} u' (a_1+a_2)$, where $u'$ is equal to
$\vec{e}_k\vec{u}$ and nonzero. Both $d_1$ and $d_2$ are
nonzero.}.
 Therefore, only one of the two eigenstates of the
$2$-dimensional subspace of $\hat{B}$ will remain an eigenstate of
$\hat{C}$. As a consequence the output has to be a mixed state,
since the `input' wave function is equal to $\vec{u}_1+\vec{u}_2$.

The same result can be derived by decomposing the operator
$\hat{B}$ into the identity matrix and one isolated element on the
diagonal. The isolated element on the diagonal times $\hat{A}$
will produce a rank-$1$ matrix. A rank-$1$ matrix is also
produced, if one switches the order of the matrices. The kernel of
each of these two rank-$1$ matrices is $(2^n-1)$-dimensional.
Therefore, except for a $2$-dimensional space, the eigenvalue and
eigenvector structure of the rest of $\hat{A}$ are unaffected  by
the multiplication with $\hat{B}$ and the symmetrization.  The
operators $\hat{C}$ and $\hat{A}$ possess the same eigenvalue and
eigenvector structure except for this $2$-dimensional subspace,
which cuts into both of the degenerate subspaces of $\hat{A}$.

The old basis is next transformed into a new basis to more easily
distinguish the pure state from the mixed state.
 We choose our new basis in such a way that the wave function  $|000\cdots00\rangle
+ |000\cdots01\rangle + \cdots +
|111\cdots10\rangle+|111\cdots11\rangle$ is mapped into the new
basis element $|000\cdots00\rangle$. The rest of the old basis can
be mapped into any new basis, as long as the transformation is
unitary.

Next, one measures the output state qubit by qubit in the new
basis. There are two possibilities. In the pure state case all the
measurements without fail produce $|0\rangle$, since one of the
eigenstates of $\hat{C}$ is exactly the `input' wave function in
the new basis $|000\cdots00\rangle$.

As mentioned, in the mixed state case only one of the eigenstates
$|000\cdots00\rangle + |000\cdots10\rangle + \cdots
+|111\cdots00\rangle+ |111\cdots10\rangle$ and
$|000\cdots01\rangle + |000\cdots11\rangle + \cdots
+|111\cdots01\rangle+ |111\cdots11\rangle$ that form a basis of
the first degenerate subspace of $\hat{A}$ remains an eigenstate
of $\hat{C}$. The transition probability of the `input' wave
function is $1/2$ to both of these basis elements.
Therefore, the separate measurements of all the qubits  in the new
basis results in at least one of the $n$ qubits to be $|1\rangle$
with probability of not less than $1/2$. Let us explain this in
more detail. We know that one of the eigenstates has a transition
probability of exactly $1/2$ to the `input' wave function. As a
consequence the rest of the eigenstates have the same transition
probability as a sum. The probability must be at most $1/2$ that
the new basis element $|000\cdots00\rangle$ will be the outcome of
the measurements and all the qubits are measured to be
$|0\rangle$. If we repeat the cycle $m$ times, the error
probability is bounded above by $2^{-m}$.

 The measurement of each qubit is carried out with the help of a
 set of $n$
properly aligned Stern-Gerlach apparatus. Each qubit of the whole
wave function will be measured separately. Such a splitting is
achieved
without damaging the coherence of the wave function (see, e.g.,
Feynman {\it et al.} \cite{fey} for an introductory discussion of
Stern-Gerlach experiments).

Alternatively, if horizontally and vertically polarized photons
are used, then the separation can be made by a polarising beam
splitter with different reflection probabilities for horizontally
and vertically polarisaed photons; for example, transmitting
horizontally polarised and reflecting vertically polarised
photons.

In the remaining paragraphs of this section we conclude the
description of the algorithm.
 After completing one cycle, one moves on
to the next cycle.
 In total $\log_2 N$ cycles have to be performed. As one moves
 from cycle to cycle the number of qubits
 needed to enumerate the remaining states decreases one by one.
Once the last cycle is finished, one has established within
certain error bounds where the {\sl marked} record can be found.

Next, we will demonstrate that the accumulated error probability,
inherent in moving from cycle to cycle and choosing smaller and
smaller subsets, can be made sufficiently {\sl small}. To show
this, define the cumulative probability of not choosing an
incorrect
 subset in any one of the cycles to be
$1-\epsilon=(1-\epsilon_{N})
(1-\epsilon_{N/2})(1-\epsilon_{N/4})\cdots (1-\epsilon_{2})$,
where each of the $\epsilon_i$ corresponds to the error
probability in one cycle
of the algorithm starting out with a fixed number of states, i.e.,
$N$, $N/2$,$\cdots$. It is easiest, if one chooses a large enough
number of iterations, e.g. $m$,
for each cycle so that the individual cycle error, decreasing
exponentially with $m$, i.e. $2^{-m+1}$, is sufficiently {\sl
small}.
This allows us to state a sequence of inequalities connecting the
cumulative error probability with the sum of the error
probabilities for each cycle, and then with the error probability
for the first cycle, which in itself is bounded above, i.e.
$\epsilon\leq \epsilon_{N}+\epsilon_{N/2}+\cdots+\epsilon_2 \leq
\log_2(N) \epsilon_{N}\leq \log_2(N) 2^{-m+1}$. This bound is
sufficiently {\sl small} for reasonable $m$\footnote{If, for
example, $m$ is chosen to be $\log_2 (N)+2$ with $\log_2 N\geq 1$,
then the cumulative error probability is always less then $1/3$.}.

Normalization factors can be introduced in the appropriate places,
but would change nothing in the basic structure of the algorithm,
since relative probabilities between different components of the
wave function are left unchanged.


\section{Conclusion} \label{sec:5}

In this section  some concluding remarks are added to round of the
paper. A number of issues can be raised in connection with the
earlier sections:

$\diamond$ Is the L\"uders postulate correct\footnote{It could be
that instead of the projection of the state onto a point in the
degenerate subspace, one has a projection onto all the points in
the degenerate subspace with an appropriate measure. This is not
completely implausible to the author.
}, i.e. to what extent has it been verified experimentally?

$\diamond$  What observables can one construct\footnote{ Besides
the Wigner-Araki-Yanase theorem, what other restrictions are there
on observables? How does one realize observables with degenerate
spectrum? Are there construction mechanism, which are
self-stabilizing?}? How does one construct them?

$\diamond$ Are the standard rules of non-relativistic quantum
mechanics applicable up to the accuracy required for the
implementation of the algorithm? There have been numerous
speculations on the limits of the rules of quantum mechanics.
Maybe this paper, together with the varied developments of quantum
computing and information theory, will give added impetus to study
these issues experimentally.

$\diamond$ In many versions of the stochastic quantum mechanics
approach there is a natural `collapse' of any input wave function
into the energy eigenstates of the Hamiltonian. {\it Identifying
the Hamiltonian with the observable $\hat{C}$ is arguably the most
promising approach for implementing the algorithm.} What evidence
is there for stochastic quantum mechanics?

 The case of having more than one {\sl marked} state, in itself
an interesting problem, is here only briefly commented on.
Naturally, the algorithm can be modified to handle an unstructured database
with a number of {\sl marked} states
\footnote{An almost identical algorithm, to the one described in
this paper, can find  a marked record in a database with
potentially a multiple number of {\sl marked} records. Again the
goal is to reduce, cycle by cycle, the subset of states to be
considered, and eventually find at least one of the {\sl marked}
states. To be able to do this, one only has to remodel slightly
Section~\ref{sec:4} of the present algorithm. The essential fact
that the degeneracy of the smaller subspace of $\hat{A}$ is broken
in $\hat{C}$ remains
unchanged.  
}. This extension has direct applications to NP-complete problems.

The purpose of this paper was to show that  the L\"uders postulate
has interesting consequences for quantum algorithms,
 and to sketch an implementation.
Issues related to implementability
 and computational complexity,
i.e. what does a polynomial number of steps mean, will be
discussed in a more detailed subsequent
paper.

The author wishes to express his gratitude to Dorje Brody for
stimulating discussions.
\newline
Electronic mail: b.meister@imperial.ac.uk

\end{document}